\documentclass[12pt]{article}
\usepackage{amssymb}
\usepackage{color,graphicx}
\usepackage{amsmath}

\newcommand{\psid}{\psi^{\dagger}}

\begin{document}

\title{Critical Behavior of Percolation Process Influenced by Random Velocity Field: One--Loop Approximation}

\author{M. Dan\v{c}o$^{1}$, M. Hnati\v{c}$^{1,2}$, T. Lu\v{c}ivjansk\'{y}$^{1,2}$, L. Mi\v{z}i\v{s}in$^{2}$
\\
$^1$ Institute of Experimental Physics, SAS, Watsonova 47, 040 01 Ko\v{s}ice \\
$^2$ Faculty of Science, P. J. \v{S}af\'{a}rik University, \v{S}rob\'{a}rova 2, 041 54 Ko\v{s}ice}

\maketitle
\begin{abstract}
Using perturbative renormalization group we investigate the influence of random velocity field on the critical behavior of directed bond percolation process near its second--order phase transition between absorbing and active phase. Antonov-Kraichnan model with finite correlation time is used for description of advecting velocity field. The field-theoretic renormalization group approach is applied for getting information about asymptotic large scale behavior
 of the model under consideration. The model is analyzed near its critical dimension through three-parameter expansion in $\epsilon,\delta,\eta$, where $\epsilon$ is the deviation from the Kolmogorov scaling, $\delta$ is the deviation from the critical space dimension $d_c$ and $\eta$ is the deviation from the parabolic dispersion law for the velocity correlator. Fixed points with corresponding regions of stability are determined to the leading order in the perturbation scheme.
\end{abstract}

\section{Introduction}
Percolation processes are  the famous models for description of random structures \cite{Jansen_Tauber_2004,Tauber_2003}. Part of these processes are known as directed bond percolation (DP). The general feature of DP is that the agent (particle) can propagate from one site to another site  in the allowed passage direction. Direction of agents spreading is designated by preferred direction of space. The DP can also serve for explaining hadron interaction at very high energy (Reggeon field theory) \cite{Reggeon}, various models of spreading disease \cite{Jansen_Tauber_2004, Tauber_2003}, stochastic reaction-diffusion processes on a lattice \cite{Hinrichsen} or as in original formulation \cite{Broad57} wetting of porous material or exploring path in labyrinth.  The upper critical dimension for this problem was estimated to be $d=4$ in contrast to the value $d=6$ for the isotropic dynamical case \cite{Jansen_Tauber_2004,Jan85,Frey94}.

One of the most important property of DP is an exhibition of non-equilibrium second-order phase transition between the absorbing and the active phase. The absorbing (inactive) phase corresponds to the case, when medium does not contain agents (sick individuals) and the active phase represents behavior of system, when the number of agents fluctuates around constant value. 

 A lot of effort was put into the investigation of various  effects, e.g.
 in papers \cite{Jan99_Hin07_Jan08} long-range interactions by the means of Levy-flight jumps were studied both in time and space variables, introduction of immunization was examined in \cite{Car85}, effect of surfaces was studied in \cite{Jan88} etc. One can also easily imagine that spreading of disease can be rapidly enhanced by some external atmospheric current or by flying insects. In both cases additional drift can be modelled by the means of random velocity field \cite{Antonov_Kapusin_2008} with prescribed
 statistical properties. In this paper the influence of advective field described by rapid-change Kraichnan model was studied, which is characterised by white-in-time nature of velocity correlator. Generalizing this approach it is possible to study e.g. effect of compressibility \cite{Ant10} or using stochastic Navier-Stokes equations effects of "real" turbulent field \cite{Ant11}. In this work we investigate the influence of finite correlated velocity field (for introduction see  \cite{Antonov_1999}) and determine how it can change the critical behavior of percolation process. By the means of renormalization
group approach we determine possible fixed points with corresponding regions of stability. We show that  the model exhibit 
$12$ possible large-scale regimes and the influence of velocity fluctuations is illustrated on quantities as the number of particles, the survival probability of an active cluster and the radius of gyration of the active particles.

The paper is organized as follows. In the section $2$ we give the brief description of the model. In the section $3$  multiplicative renormalizability of the model  is demonstrated and the RG functions ($\beta$ functions and anomalous dimensions) are calculated to the one-loop approximation, in the section $4$ possible large-scale regimes are listed and their physical meaning  is given in the section 5.
\section{Field-Theoretic Formulation of the Model}
Two different approaches for field-theoretic formulation of DP  are possible. The first approach is more rigorous and is based on the use of master equation, that can be rewritten employing Doi formalism \cite{Doi} into the form of time--dependent Schr\"{o}dinger equation with non--hermitian Hamiltonian. After the continuum limit is performed the effective action is obtained, which is amenable to the usual field--theoretic methods. Action for the pure DP problem \cite{Jansen_Tauber_2004} can be written in the following form
\begin{eqnarray}
S_{1} &=& \psid (-\partial_t + D_0 \nabla^2 - D_0 \tau_0) \psi +\frac{D_0 \lambda_0}{2} [ (\psid)^2 \psi - \psid \psi^2 ],
\label{eq:action_DP}
\end{eqnarray}
where all required  integrations over space-time variables are implied. 
Here $\psi\equiv\psi(t,{\mathbf x})$ is the coarse-grained density of infected individuals (agents), $\psid$ is the response function, $D_0$ is the diffusion constant, $\lambda_0$ is positive coupling constant and $\tau_0$ is deviation from the threshold value of the injected probability.  One can assume $\tau_0\sim p_c-p$, where $p_c$ is a 
 critical probability for observing percolation (analogous to the deviation from critical temperature for
 equilibrium models). 
It can be readily shown \cite{Jansen_Tauber_2004} that the second approach based on the use of Langevin equation leads after appropriate rescaling to the same coarse--grained action functional (\ref{eq:action_DP}). The both of them lead to the same predictions for the universal quantities.

The agents can be considered as passive scalar quantity \cite{Antonov_1999} that is advected by the velocity field with no back influence on the
velocity field itself with nontrivial interactions given by the cubic terms in (\ref{eq:action_DP}). 
The inclusion of the velocity field ${\mathbf v}(t,{\mathbf x})$ corresponds to the replacement partial derivation with respect to time by convective derivation
$\nabla_t = \partial_t +(\mathbf{v}\cdot \nabla)$ in equation (\ref{eq:action_DP}).  In contrast to the work of Antonov et al. \cite{Antonov_Kapusin_2008} we assume that the velocity field is a random Gaussian variable with zero mean and correlator in the following form \cite{Antonov_1999}
\begin{equation}
\langle v_i(t,\mathbf{x}) v_j (t',\mathbf{x'}) \rangle = 
\int \frac{\textrm{d} \omega}{2\pi} \int \frac{\textrm{d}\mathbf{k}}{(2\pi)^d}P_{ij}^{k} 
D_v (\omega ,k) {\mathrm e}^{-i\omega(t-t')+ i\mathbf{k}(\mathbf{x}-\mathbf{x}')},
\label{eq:correlator_velocity_field}
\end{equation}
where $P_{ij}(k) = \delta_{ij} -k_i k_j /k^2$ is transverse projection operator and $D_v$ has the following form \cite{Antonov_1999}
\begin{equation}
D_v (\omega,k) = \frac{g_{10} D_0^3 k^{4-d-2\epsilon -\eta}}{\omega^2 + u_{10}^2 D_0^2 (k^{2-\eta})^2}  ,
\label{eq:kernel_function}
\end{equation}
where $g_{10}$ is the coupling constant and $\epsilon$, $\eta$ play the role of small expansion parameters. In this paper $\epsilon$ should be understood as deviation from Kolmogorov scaling \cite{Frisch} and $\delta$ is the deviation from the space dimension four via relation $d = 4 - 2\delta$. The exponent $\eta$ are related to the frequency $\omega \propto k^{2-\eta}$ and $\eta=4/3$ corresponds to the Kolmogorov frequency.  The averaging procedure with respect to the velocity fluctuations may be performed with the use of Gaussian-like action functional
$S_{2} = -\frac{1}{2} \mathbf{ v} D^{-1}_v  \mathbf{v}$.

The considered model (\ref{eq:correlator_velocity_field}), (\ref{eq:kernel_function})  contains two important limits. 
The first of them is known as rapid--change model \cite{Antonov_1999}, that is defined by
 $u_{10} \rightarrow \infty$ a $g_{10}' \equiv g_{10}/u_{10}^2 = $ const.,
$D_v (\omega,{\mathbf k}) \rightarrow g_{10}' D_0  k^{-d-2\epsilon +\eta}$.
Hence we see, that velocity correlator is decorrelated (white noise) in time variable. The second limit is called 'frozen' velocity field \cite{Antonov_1999} and is obtained as $u_{10} \rightarrow 0$ and $g_0''=g_0/u_{10}=$ const., 
$D_v (\omega,{\mathbf k}) \rightarrow g_{10}'' D_0^2  k^{-d+2-2\epsilon} \pi \delta(\omega)$.
In this case the velocity correlator is independent of time in the $t$-representation. 

The full problem is equivalent to the sum of functionals corresponding to the DP and velocity field.
\begin{equation}
S_0 = \psid [-\partial_t - (\mathbf{v}\cdot\nabla) + D_0 \nabla^2 - D_0 \tau_0] \psi
+\frac{D_0 \lambda_0}{2} [ (\psid)^2 \psi - \psid \psi^2 ] -\frac{1}{2} \mathbf{ v} D^{-1}_v  \mathbf{v}.
\label{eq:total_action}
\end{equation}
In perturbation theory it can be easily shown \cite{Antonov_Kapusin_2008}, that expansion parameter is rather $\lambda_0^2$ than $\lambda_0$, which
is a consequence of the so-called rapidity reversal symmetry (in the language of Reggeon field theory)
\begin{equation}
  \psi(t,{\mathbf x}) \rightarrow -\psi^\dagger(-t,{\mathbf x}),\quad
    \psi^\dagger(t,{\mathbf x}) \rightarrow -\psi(-t,{\mathbf x}).
\label{eq:dp_symmetry}
\end{equation}
 Therefore the introduction of the following new charge $g_{20} = \lambda_0^2$ is appropriate.

\section{Canonical Dimensions and UV divergences}

Theoretical analysis of UV divergences is based on the standard power counting \cite{Vasiliev,Zinn-Justin}. Dynamic models 
involve two independent scales: the time (frequency) and length (momentum) scale. Therefore the 
canonical dimension of any quantity $F$ is determined by two dimensions: the frequency dimension 
$d_F^{\omega}$ and the momentum dimension $d_F^k$. These dimensions are found out from the 
usual normalization condition
\begin{equation}
d_{\omega}^{\omega} = - d_t^{\omega} = 1, \hspace{0.5cm}
d_{k}^{k} = - d_k^{x} = 1, \hspace{0.5cm}
d_{k}^{\omega} = - d_{\omega}^k = 0
\end{equation}
and from the requirement that the action (\ref{eq:total_action}) should be dimensionless (with respect to the momentum and frequency dimensions separately). 
The total canonical dimension is given by the relation $d_F =d_F^k+2d_F^{\omega}$ 
(in order to have parabolic dispersion law $\partial_t \propto \nabla^2$ for the free theory). 
It plays the same role in the theory of renormalization  as length (momentum) dimension for static 
theory \cite{Vasiliev}. The canonical dimension of the model (\ref{eq:total_action}) are given
 in the Table \ref{tab:canonical_dimension}. The model is logarithmic at space dimension $d=4$ (or $\delta = 0$) and 
 for $\epsilon = \eta = 0$. The  UV divergences in the minimal subtraction (MS) scheme
  realize themselves as poles in $\epsilon$, $\delta$, $\eta$ or their linear combination.
\begin{table}[!ht]
\begin{center}
\caption{Canonical dimensions of fields and bare parameters of  action (\ref{eq:total_action}).}
\begin{tabular}{c @{\hspace{0.5cm}} c  @{\hspace{0.5cm}} c  @{\hspace{0.5cm}} c  @{\hspace{0.5cm}} c  @{\hspace{0.5cm}} c  @{\hspace{0.5cm}} c  @{\hspace{0.5cm}} c  @{\hspace{0.5cm}} c @{\hspace{0.5cm}} c @{\hspace{0.5cm}} c}
\hline
$F$ & $\psi$ & $\psi^{\dagger}$ & ${\bf v}$ & $D_0$ & $\tau_0$ & $\lambda_0$ & $g_{10}$ & $g_{20}$ & $u_{10}$ & $u_{20}$ \\
\hline 
\hline
$d_F^k$ & $d/2$ & $d/2$ & $-1$ & $-2$ & $2$ & $\delta$ & $2\epsilon + \eta $ & $2\delta$ & $\eta$ & $0$ \\
\hline
$d_F^{\omega}$ & $0$ & $0$ & $1$ & $1$ & $0$ & $0$ & $0$ & $0$ & $0$ & $0$\\
\hline
$d_F$ & $d/2$ & $d/2$ & $1$ & $0$ & $2$ & $\delta$ & $2\epsilon+ \eta$ & $2\delta$ & $\eta$ & $0$ \\
\hline
\hline
\end{tabular}
\label{tab:canonical_dimension} 
\end{center}
\end{table}
The total canonical dimension $d_{\Gamma_N}$ of an arbitrary 1-irreducible Green function $ \Gamma_N \equiv \langle \Phi \dots \Phi \rangle$ is given by the following relation \cite{Vasiliev, Zinn-Justin}:
$d_{\Gamma_N} = d + 2 - N_{\Phi}d_{\Phi}$,
where $N_{\phi} = \{ N_v, N_{\psi}, N_{\psi^{\dagger}}\}$ are the number of fields entering into the Green function $\Gamma_N$, including summation over all types of fields $\Phi$. In logarithmic theory superficial UV divergences can affect only those functions $\Gamma_N$, for which $d_{\Gamma_N}$ is a non-negative integer. Green functions must contain the both fields $\psi$ and $\psi^{\dagger}$ because function $\langle \psi \dots \psi \rangle$ or $\langle \psi^{\dagger} \dots \psi^{\dagger} \rangle$  includes closed loop of retarded propagator \cite{Antonov_Kapusin_2008,Vasiliev}. Therefore superficial UV divergences can be present only in the functions:
\begin{eqnarray*}
& &\langle  \psi^{\dagger} \psi \rangle \hspace{0.5cm} \textrm{with the counterterms} \hspace{0.5cm} \psi^{\dagger} \partial_t \psi, \psid \partial^2 \psi, \psid \psi, \nonumber \\
& &\langle  \psi^{\dagger} \psi \psi \rangle \hspace{0.5cm} \textrm{with the counterterm} \hspace{0.5cm} \psi^{\dagger}  \psi^2, \nonumber \\ 
& &\langle  \psi^{\dagger} \psid  \psi \rangle \hspace{0.5cm} \textrm{with the counterterm} \hspace{0.5cm} (\psi^{\dagger})^2  \psi, \nonumber \\
& &\langle  \psi^{\dagger} \psi v \rangle \hspace{0.5cm} \textrm{with the counterterm} \hspace{0.5cm} \psi^{\dagger} (v\partial) \psi, \nonumber\\
& &\langle \psid \psi v v \rangle \hspace{0.5cm} \textrm{with the counterterm} \hspace{0.5cm} \psi^{\dagger} \psi v v. 
\end{eqnarray*} 
The terms $(\psid)^2 \psi$ and $\psid \psi^2$ can be renormalized by the same renormalization constant
as a consequence of symmetry (\ref{eq:dp_symmetry}) and counterterm produced by the Green function $\langle \psid \psi v \rangle$ reduces to the form $\psid (v \partial) \psi = -\psi (v \partial) \psid$ owing to the transversality of velocity field. 
In Kraichnan (rapid change) model the counterterm $\psi^{\dagger} \psi v v$ is absent and renormalization constants for
terms $\psi^{\dagger} \partial_t \psi$ and $\psi^{\dagger} (v\partial) \psi$  are the same (convective derivation 
 $\nabla_t = \partial_t +(\mathbf{v}\cdot \nabla)$ conserves its form during renormalization) due to Galilean invariance of the model. 
  We must stress that our model is not Galilean invariant due to the form of velocity correlator (\ref{eq:kernel_function}) which reflects
the existence of  sweepping effect  in developed turbulence \cite{Vasiliev}. 
As a consequence the renormalization constants at $\psi^{\dagger} \partial_t \psi$ and $\psi^{\dagger} (v\partial) \psi$  terms are different in this 
case. Moreover, to assure multiplicative renormalization of the model we have to add a new term 
$\psi^{\dagger} \psi v v$ into the action (\ref{eq:total_action}) with new independent parameter (charge) $u_{2}$ which will compensate 
divergencies produced by the correlation function 
$\langle \psid \psi v v \rangle $. The resulting renormalized action can be written in the following form 
\begin{eqnarray}
S_R &=& \psid (-Z_1\partial_t - Z_2(\mathbf{v}.\nabla) + Z_3 D \nabla^2 - Z_4 D \tau) \psi 
 \nonumber \\ &+& Z_5 \frac{D \lambda}{2} \Big[ (\psi^+)^2 \psi - \psi^+ \psi^2 \Big] 
 +Z_6 \frac{u_2}{2D} \psid \psi \mathbf{v}^2  - \frac{1}{2} \mathbf{ v} D^{-1}_v  \mathbf{v}. 
\label{eq:renor_action}
\end{eqnarray}
We note that due to the dimensional reasons the charge $u_2$ must appear combination $u_2/D$, otherwise
it would possess nonzero dimension with respect to the spatial and time variable respectively.
On the other hand the renormalized action can be obtained by the multiplicative renormalization of the fields $\psi \rightarrow \psi Z_{\psi}$, $\psi^{\dagger} \rightarrow \psi^{\dagger} Z_{\psi^{\dagger}}$ and $v \rightarrow v Z_v$ and parameters:
\begin{align}
&D_0 = D Z_D, \hspace{0.5cm} g_{10} = g_1 \mu^{2\epsilon+\eta} Z_{g_1},\hspace{0.5cm} \tau_0 = \tau Z_{\tau}, \hspace{0.5cm}  
u_{10} =u_1 \mu^{\eta} Z_{u_1}, \nonumber \\
&\lambda_0 = \lambda \mu^{\delta} Z_{\lambda}, \hspace{0.4cm} g_{20} = g_2 \mu^{2\delta} Z_{g_2}, \hspace{0.65cm} u_{20} = u_2 Z_{u_2}.
\label{eq:renorm_parameter}
\end{align}
The renormalization constant of fields and parameters (\ref{eq:renorm_parameter}) are related with renormalization constant of the action  (\ref{eq:renor_action}) through the following relations
\begin{align}
& Z_v =  Z_1^{-1} Z_2, \hspace{1.2cm} Z_{\psi}=Z_{\psi^{\dagger}}=Z_1^{1/2}, \hspace{1.2cm} Z_{\lambda}=Z_1^{-1/2}  Z_3^{-1} Z_5 , \nonumber \\ 
& Z_D = Z_1^{-1} Z_3, \hspace{1cm} Z_{u_1} = Z_1 Z_3^{-1}, \hspace{2cm} Z_{\tau} =Z_3^{-1} Z_4, \nonumber \\ 
& Z_{g_1} = Z_1^5 Z_2^{-2} Z_3^{-3},\hskip0.25cm  Z_{g_2} = Z_1^{-1} Z_3^{-2} Z_5^2, \hspace{1.1cm} Z_{u_2}= Z_2^{-2} Z_3 Z_6 . 
 \label{eq:renorm}
\end{align}
The standard RG perturbative approach is based on the Feynman diagrammatic technique \cite{Vasiliev,Zinn-Justin}.
Explicit calculation in the one--loop approximation leads to following results for renormalization constants
\begin{align}
&Z_1 = 1 + \frac{g_2}{8\delta}, \hspace{1cm} 
 Z_2 = 1 + \frac{g_2}{8\delta}, \hspace{1cm} 
 Z_3 = 1 + \frac{g_2}{16\delta} - \frac{3 g_1}{8u (1+u) \epsilon}, \nonumber \\
&Z_4 = 1 + \frac{g_2}{4 \delta}, \hspace{1cm}
 Z_5 = 1 + \frac{g_2}{2\delta}, \hspace{1cm}
 Z_6 = 1 + \frac{g_2}{16 \delta} - \frac{3 g_1(1+u_2)}{8 u_1 (1+u_1)\epsilon} .
  \label{ren_con}
\end{align}
 The basic RG differential equation for the renormalized Green function $G_R$ is given by the equation
$\{ D_{RG} + N_{\psi} \gamma_{\psi} + N_{\psid} \gamma_{\psid} \} G_R(e, \mu, \dots)=0,$
where $e$ is the full set of renormalized counterparts of the bare parameters $e_0 =\{D_0, \tau_0, u_{10},u_{20}, g_{10}, g_{20} \}$
and $\dots$  denotes other parameters, such as spatial or time variables.
 The RG operator $D_{RG}$ can be written in the form
\begin{equation}
D_{RG}  = \mu \partial_{\mu} + \beta_{u_1} \partial_{u_1} + \beta_{u_2} \partial_{u_2}+
\beta_{g_1} \partial_{g_1} + \beta_{g_2} \partial_{g_2} - 
\gamma_D D_D - \gamma_{\tau} D_{\tau},
\end{equation}
where $D_x = x \partial_x$ for any variable $x$, $\gamma_a = \tilde{D}_{\mu} \ln  Z_a$ is an anomalous dimension and $\tilde{D}_{\mu}$ denotes the differential operator taken at fixed values of the bare parameters. The $\beta$ function is defined as $\beta_g = \tilde{D}_{\mu} g, g\in\{ g_1, g_2, u_1, u_2\}$ and  from the relations (\ref{eq:renorm_parameter}) the following expressions can be obtained in straightforward manner
\begin{eqnarray*}
& & \beta_{g_1} = g_1 (-2\epsilon - \eta + 3\gamma_D), \qquad
 \beta_{g_2} = g_2 (-2\delta -\gamma_{g_2}),  \\
& & \beta_{u_1} = u_1(-\eta +\gamma_D), \hskip1.85cm
\beta_{u_2} = - u_2 \gamma_{u_2}.
\label{eq:beta_functions}
\end{eqnarray*}
From the explicit results (\ref{ren_con}) and relations (\ref{eq:renorm}) the needed anomalous dimensions 
are easily calculated
\begin{eqnarray}
 & & \gamma_D = \frac{3 g_1}{4 u_1 (1+u_1)} +\frac{g_2}{8}, \quad
\gamma_{g_2} = - \frac{3g_1}{2u_1(1+u_1)}-\frac{3g_2}{2}, \nonumber\\
& & \gamma_{u_2} = - \frac{g_2}{8} + \frac{3g_1 (1+u_2)}{4u_1(1+u_1)}.
 \label{eq:anomal}
\end{eqnarray}

\section{Fixed Points and Scaling Regimes}

According to the renomalization group theory, the infrared (IR) asymptotic behavior is governed by IR attractive fixed points (FPs).
 The fixed points $g^{*} =\{g_1^{*}, g_2^{*}, u_1^{*},u_2^{*} \} $ can be found from the requirement that all $\beta$ functions simultaneously vanish $\beta_{g_1} (g^{*}) = \beta_{g_2} (g^{*}) =\beta_{u_1} (g^{*})= \beta_{u_2} (g^{*})  = 0$.
The type of FP is determined by the eigenvalues of the matrix  $\Omega =\{ \Omega_{ij} = \partial \beta_i / \partial g_j \}$, where $\beta_i$ is the full set of $\beta$ functions (\ref{eq:beta_functions}) and $g_j$ is the full set of charges $\{g_1, g_2, u_1,u_2 \}$. 
For the IR attractive FP the eigenvalues of the matrix $\Omega$ are strictly positive quantities. From this condition the region of stability for
 the given FP can be determined.
It is well known that in IR asymptote the Green functions exhibit scaling behavior. The critical dimension $\Delta_F$ of the IR quantity is determined by the following relation
$\Delta_F = d_F^k +\Delta_{\omega} d_F^{\omega} + \gamma_F^{*}$,
where $d_F^{k, \omega}$ are canonical dimension of quantity $F$ from Tab. \ref{tab:canonical_dimension}, $\gamma_F^{*}$ is value of the anomalous dimension at the given fixed point and $\Delta_{\omega} = 2 - \gamma_D^{*}$. 

Now we introduce the quantities \cite{Jansen_Tauber_2004}, that could be used for illumination
of the fixed points' structure of the model. 
 The first quantity is the number of active particles $N(t)$, which are spreading from origin $\mathbf{x} = 0$ at time $t = 0$ and is given by the response function  as follows
\begin{equation}
N(t)=\int \textrm{d} \mathbf{x} \hspace{2pt} \langle \psi(t, \mathbf{x} ) \psid(0,{ \bf 0} )\rangle \propto t^{\frac{d - 2\Delta_{\psi}}{\Delta_{\omega}}} = t^{ 
\frac{ g_2^{*} }{ 4( 2 - \gamma_D^{*}) }},
\label{eq:Nt}
\end{equation}
where $\gamma_D^{*} = \gamma_D (g_1^{*}, g_2^{*}, u^{*})$. The second quantity is 
the survival probability $P(t)$ of an active cluster and is defined by the following expression
\begin{equation}
P(t) = - \lim\limits_{A\rightarrow \infty} \langle \mathrm{e}^{-A\psi(0,\mathbf{x})} \psid(-t, \mathbf{0} )  \rangle \propto t^{-\frac{\Delta_{\psi}}{\Delta_{\omega}}} = t^{-\frac{2-\delta-\frac{g_2^{*}}{8}}{2-\gamma_D^{*}}}.
\label{eq:Pt}
\end{equation}
The radius of gyration $R(t)$ of the active particles is defined as follows
\begin{equation}
R^2(t) = \frac{\int \mathrm{d} \mathbf{x} \hspace{2mm} \mathbf{x}^2 \langle \psi (t, \mathbf{x}) \psid (0,\mathbf{0}) \rangle}{\int \mathrm{d} \mathbf{x} \hspace{2mm} \langle \psi (t, \mathbf{x}) \psid (0,\mathbf{0}) \rangle } \propto t^{\frac{2}{\Delta_{\omega}}} = t^{\frac{2}{2-\gamma_D^{*}}}
\label{eq:R2t}
\end{equation}
In order to  investigate the FPs of this model, we use the exact relation for beta
 function $\beta_{g_1}/g_1-3\beta_{u_1}/u_1=2(\eta-\epsilon)$. It follows that beta functions $\beta_{g_1}$ and $\beta_{u_1}$  can vanish simultaneously for finite value of their arguments only in the case $\epsilon=\eta$, which  should be studied separately. On the other hand, when $\epsilon \neq \eta$ it is necessary to consider limiting behavior, i.e. either $u_1=0$ or $u_1 =\infty$ and rescale $g_1$ so that in beta functions $\beta_{g_1}$, $\beta_{u_1}$ anomalous dimension $\gamma_D$ acquires finite value.
\begin{table}[ht!]
\begin{center}
\caption{Stable FPs of the "frozen" velocity field with corresponding critical exponents.}
\renewcommand{\arraystretch}{1.5}
\begin{tabular}{|c|c|c|c|c|c|}
\hline
FP & FP 1 &  FP 2 &  FP 3 & FP 4a & FP 4b\\
\hline
\hline
${g_1^{*\prime\prime}}$ & 0 & 0 & $ \frac{4 \epsilon}{3}$ & $\frac{ 4(6\epsilon-\delta)}{15}$ & $\frac{ 4(6\epsilon-\delta)}{15}$\\
$g_2^{*}$ & 0 & $ \frac{4 \delta }{3}$ & 0 & $ \frac{8(\delta-\epsilon)}{5}$ &  $ \frac{8(\delta-\epsilon)}{5}$\\
$u_1^{*}$ & 0 & 0 & 0 & 0 & 0 \\
$u_2^{*}$ & not fixed & $0$ & $-1$  & $0$ & $\frac{2\delta-7\epsilon}{6\epsilon-\delta}$\\
\hline
\hline
Region & $\epsilon<0$ & $6 \epsilon < \delta$ & $\epsilon>0$ & $0< \epsilon $ & $0< \epsilon $\\
of stability & $\delta<0$ &  $\delta>0$ &  $\epsilon>\delta$ & $\frac{7}{2}\epsilon<\delta<6\epsilon$ & $\epsilon<\delta<\frac{7}{2}\epsilon$ \\
 & $\eta<0$ & $6\eta<\delta$ & $\eta<\epsilon$  & $\eta< \epsilon$ &$\eta< \epsilon$\\
\hline
\hline
$\Delta_{\omega}$ & $2$ & $2-\frac{\delta}{6}$ & $2-\epsilon$ & $ 2-\epsilon $ &$ 2-\epsilon $ \\
$- \Delta_{\psi} / \Delta_{\omega}$ & $ \frac{\delta - 2}{2}$ & $ \frac{7 \delta -12}{12-\delta}$ & 
$\frac{\delta -2}{2-\epsilon}$ & $ \frac{6 \delta - \epsilon - 10}{5(2-\epsilon)}$ &
 $ \frac{6 \delta - \epsilon - 10}{5(2-\epsilon)}$ \\
$(d - 2 \Delta_{\psi} )/ \Delta_{\omega}$ & $0$ & $\frac{2\delta}{12 - \delta}$ & $0$ & $\frac{2}{5}\frac{\delta-\epsilon}{2-\epsilon} $ & 
$\frac{2}{5}\frac{\delta-\epsilon}{2-\epsilon} $\\
\hline
\end{tabular}
\label{tab:frozen}
\end{center}
\end{table}
Let us consider case, for which limit of correlator velocity field is known as a 'frozen' velocity field and exhibits four FPs Tab. \ref{tab:frozen}. In this regime  the new variable $g_1^{\prime\prime} \equiv g_1/u_1 $ was introduced and its beta function has the form  $\beta_{g_1^{\prime\prime}} = g_1^{\prime\prime} (-2 \epsilon +2 \gamma_D)$.
The FP 1 correspond to the Gaussian (free) fixed point. In this regime the radius of gyration $R(t)$
 imitates the ordinary random walk ($R (t) \propto t^{1/2}$) behavior. For the FP 2 the correlator of
 the velocity field is irrelevant and model behaves like the "pure" DP class with corresponding 
  critical exponents \cite{Jansen_Tauber_2004}. For the FP 3, the percolation nonlinearity of the DP action (\ref{eq:action_DP}) is irrelevant. Finally, the FP 4a and FP 4b corresponds to the nontrivial IR scaling regime, in which the nonlinearity of DP model (\ref{eq:action_DP}) and velocity field are both important. Moreover, for the FP 4a  the 
  interaction $\psid \psi \mathbf{v}^2$ is irrelevant unlike for the FP 4b. In the interval $\epsilon \in (0,2)$
  and for $\delta= $const. the exponent of survival probability $P(t)$ is less than $-1$ for the FP 3 and greater then $-1$ for FP 4. It follows that in regime  FP 4 the survival probability decrease more slowly than for FP 3. In the regime FP 3 the velocity fluctuations leads to the decay of an active cluster, but in regime FP 4 leads to the oposite effect.
\begin{figure}[ht]    
\begin{center}     
\includegraphics[height=40mm]{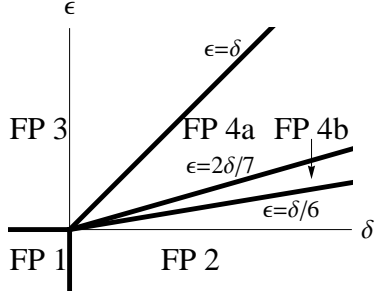}  
\end{center}   
\vspace{-2mm} \caption{Region of stability for fixed points of frozen velocity field.}
\label{fig:region}
\end{figure}  
   In the Fig. \ref{fig:region} the regions of stability for fixed point in the plane $\epsilon-\delta$ for $\eta \leq 0$ are depicted. The boundaries of the regions are represented by the thick lines.

The second limit case is characterized by white-in-time nature of velocity field. This regime is  called the rapid change model \cite{Antonov_1999} and exhibits four FPs. For this case it is advantageous to introduce new variables
 $g_1^{\prime}\equiv g_1/u_1^2$ and $w = 1/u_1$, for which corresponding beta functions have the following form $\beta_{g_1^{\prime}} = g_1^{\prime} (\eta-2 \epsilon + \gamma_D)$, $\beta_w = w(\eta-\gamma_D)$. The FPs and their regions of stability were published in the work Antonov et al. \cite{Antonov_Kapusin_2008}.
\begin{table}[ht]
\begin{center}
\caption{Fixed points for the non-trivial case.}
\renewcommand{\arraystretch}{1.5}
\begin{tabular}{|c|c|c|c|}
\hline
FP & FP 5 & FP 6a & FP 6b\\
\hline
\hline
$\frac{g_1^{*}}{u_1^{*}(1+u_1^{*})}$ & $ \frac{4 \epsilon }{3}$ & $\frac{ 4(6\epsilon-\delta)}{15}$ &
$\frac{ 4(6\epsilon-\delta)}{15}$ \\
$g_2^{*}$ & $0$ & $ \frac{8(\delta-\epsilon)}{5}$ & $ \frac{8(\delta-\epsilon)}{5}$\\
$u_2^{*}$ & $-1$ & $0$ & $\frac{2\delta-7\epsilon}{6\epsilon-\delta}$ \\
\hline
\hline
Region & $\epsilon>0$ &  $\epsilon>0$ & $\epsilon>0$ \\
 of stability  &$\eta=\epsilon>\delta$ & $\eta=\epsilon<\frac{2\delta}{7}$ &$\eta=\epsilon>\frac{2\delta}{7}$ \\
\hline
\hline
$\Delta_{\omega}$ & $2-\epsilon$ & $ 2-\epsilon $ & $ 2-\epsilon $ \\
$- \Delta_{\psi} / \Delta_{\omega}$ & $\frac{\delta -2}{2-\epsilon}$ & $ \frac{6 \delta - \epsilon - 10}{5(2-\epsilon)}$ &
$ \frac{6 \delta - \epsilon - 10}{5(2-\epsilon)}$\\
$(d - 2 \Delta_{\psi} )/ \Delta_{\omega}$ & $0$ & $\frac{2}{5}\frac{\delta-\epsilon}{2-\epsilon} $ & 
$\frac{2}{5}\frac{\delta-\epsilon}{2-\epsilon} $\\
\hline
\end{tabular}
\label{tab:fp_3_case}
\end{center}
\end{table}

The third, most non-trivial case, differs from the previous ones, the beta functions $\beta_{g_1}$ and $\beta_{u_1}$ are
 proportional to each other and thus the FPs are degenerated. However, the FPs can be divided 
 according the value of $g_2$ into the two groups (see Tab. \ref{tab:fp_3_case}). For the FP $5$ the nonlinearity of the DP action (\ref{eq:action_DP}) is irrelevant. For the FP $6a$ and FP $6b$ the interaction part of DP action (\ref{eq:action_DP}) and correlator velocity field coexist. In the interval $\epsilon \in (0,2)$ the
   exponents of quantities ($N(t)$, $P(t)$, $R(t)$) behave in the same way as FP $3$, FP $4a$ and FP $4b$, respectively.
\section{Conclusion}
In this work we have investigated the influence of the finite correlated velocity field (Antonov-Kraichnan model) on the progress of directed bond percolation of process. After the construction of  the field--theoretic model, we have performed the calculations of the renormalization constants of the triple expansion in $\epsilon$, $2 \delta =  4- d$ and $\eta$ in the leading order. Fixed points, their regions of stability and behavior of the number of active particles, survival probability and radius of gyration were calculated. 
\section*{Acknowledgments}

The work was supported by VEGA grant 1/0222/13 of the Ministry 
of Education, Science, Research and Sport of the Slovak Republic, by
Centre of Excellency for Nanofluid of IEP SAS. One of the authors (Luk\'{a}\v{s} Mi\v{z}i\v{s}in) 
thanks for the support the VVGS grant of UPJS. This
article was also created by implementation of the Cooperative phenomena
and phase transitions in nanosystems with perspective utilization in nano-
and biotechnology project No 26220120033 and No 26110230061. Funding
for the operational research and development program was provided by the
European Regional Development Fund.


\begin{thebibliography}{99}

\bibitem{Jansen_Tauber_2004} H.-K.~Janssen and U. C. T\"{a}uber, Ann. Phys. {\bf 315}, 147 (2004).
\bibitem{Tauber_2003} U. C. T\"{a}uber, Adv. Solid State Phys. {\bf 43}, 659 (2003).
\bibitem{Reggeon} J. L. Cardy and R. L. Sugar, J. Phys. A Math. Gen. {\bf 13}, L423 (1980).
\bibitem{Hinrichsen} H. Hinrichsen, Adv. Phys. {\bf 49}, 815-958 (2000). 
\bibitem{Broad57} S.R.~Broadbent and I.M.~Hamersley, Proc. Cambr. Phil. Soc. {\bf 53}, 629 (1957).
\bibitem{Jan85} H.K.~Janssen, Z. Phys. B {\bf 58}, 311 (1985).
\bibitem{Frey94} E.~Frey, U.C.~T\"auber and F.~Schwabl, Phys. Rev E {\bf 49}, 5058 (1994).
\bibitem{Jan99_Hin07_Jan08} H.-K.~Janssen, K.~Oerding, F.~van~Wijland and H.J. Hilhorst, Eur. Phys. J. B {\bf 7}, 137 (1999); H.~Hinrichsen, J. Stat. Phys.: Theor. Exp. P07066 (2007); H.-K.~Janssen and O.~Stenull, Phys. Rev. E {\bf 78}, 061117 (2008);
\bibitem{Car85} J.L.~Cardy and P.~Grassberger, J. Phys. A: Math. Gen. {\bf 18} L267 (1985)
 J.L.~Cardy, J. Phys. A: Math. Gen. {\bf 16}, L709 (1983);
 F.~Linder, J. Tran-Gia, S.R.~Dahmen and H.~Hinrichsen, J. Phys. A: Math. Theor. {\bf 41}, 185005 (2008). 
\bibitem{Jan88}  H.-K.~Janssen, B.~Schaub and B.~Schittmann, Z. Phys. B {\bf 72}, 111 (1988).
\bibitem{Antonov_Kapusin_2008} N. V. Antonov, V. I. Iglovikov, A. S. Kapustin, J. Phys. A: Math. Theor. {\bf 42}, 135001 (2008).
\bibitem{Ant10} N.V.~Antonov and A.S.~Kapustin, J. Phys. A: Math. Theor. {\bf 43}, 405001 (2010).
\bibitem{Ant11} N.V.~Antonov, A.S.~Kapustin and A.V.~Malyshev,	Theor. Math. Phys. {\bf 169}, 1470-1480 (2011).
\bibitem{Antonov_1999} N. V. Antonov, Phys. Rev. E {\bf 60}, 6691 (1999); Physica D 144, 370 (2000).
\bibitem{Doi} M.~Doi, J. Phys. A. {\bf 9}, 1456 (1976); J. Phys. A {\bf 9}, 1479 (1976).
\bibitem{Frisch} U.Frisch, \textit{ Turbulence: The Legacy of A. N. Kolmogorov},  (Cambridge University Press, Cambridge, 1995).
\bibitem{Vasiliev} A. N. Vasil'ev, \textit{ The Field Theoretic Renormalization Group in Critical Behavior Theory and Stochastic Dynamics}, (Boca Raton: Chapman Hall/CRC 2004).
\bibitem{Zinn-Justin} J. Zinn-Justin, \textit{ Qauntum Field  Theory and Critical Phenomena}, (Oxford Univ. Press, Oxford, 1989). 
\end{thebibliography}
\end{document}